\documentclass[12pt]{iopart}

\usepackage{epic}
\usepackage{eepic}
\usepackage{iopams}
\usepackage{latexsym}
\usepackage{isolatin1}
\usepackage{graphicx}

\newcommand{\ket}[1]{|#1\rangle}
\newcommand{\bra}[1]{\langle#1|}
\renewcommand{\vec}{\mathbf}
\begin{document}

\title{Dynamical Structure Factors for Dimerized Spin Systems} 

\author{M Müller$^1$, H-J Mikeska$^1$, and N Cavadini$^2$}
\address{$^1$\ Institut für Theoretische Physik,Universität Hannover, 
Appelstrasse 2, D-30167 Hannover, Germany}
\address{$^2$\ Laboratory for Neutron Scattering, ETH Zürich \& Paul 
Scherrer Institut, CH-5232 Villigen PSI, Switzerland}

\begin{abstract}
  We discuss the transition strength between the disordered ground state and
  the basic low-lying triplet excitation for interacting dimer materials by
  presenting theoretical calculations and series expansions as well as
  inelastic neutron scattering results for the material $\rm KCuCl_3$. We
  describe in detail the features resulting from the presence of two
  differently oriented dimers per unit cell and show how energies and spectral
  weights of the resulting two modes are related to each other. We present
  results from the perturbation expansion in the interdimer interaction
  strength and thus demonstrate that the wave vector dependence of the simple
  dimer approximation is modified in higher orders. Explicit results are given
  in $10^{\rm th}$ order for dimers coupled in 1D, and in 2$^{\rm nd}$ order
  for dimers coupled in 3D with application to $\rm KCuCl_3$ and $\rm
  TlCuCl_3$.
\end{abstract}


\maketitle

\section{Introduction}
Low-dimensional quantum antiferromagnets have received much interest in recent
years since they serve as model substances allowing to investigate in detail
the effects of quantum fluctuations and to test theoretical models.  One
important class of materials in this context consists of an assembly of dimers
(two strongly coupled spins 1/2) which interact sufficiently weakly to
guarantee that the dimer gap does not close.  These materials are
characterized by a disordered singlet ground state and a finite spin gap to
triplet excited states. Prominent examples in this class are $\rm KCuCl_3$ and
$\rm TlCuCl_3$ which have been investigated in detail in the last years by
static and dynamic methods as well as theoretically \cite{Rice02}.  The most
detailed experimental information is obtained from inelastic neutron
scattering (INS) experiments, which directly explore the basic singlet-triplet
transition in all of reciprocal space \cite{Cavadini01/1, MMueller02}.

The energy of the singlet-triplet transition along the principal axis
in reciprocal space as measured in these experiments is well described
by the model of interacting dimers; to lowest order this is formulated
as effective dimer model \cite{Cavadini00/1, MMueller00}, 
and it has been
refined by perturbative cluster expansions up to $6^{\rm th}$ order
\cite{MMueller02}. Here we supplement this analysis by discussing the
dynamical structure factor. 

The dynamical structure factor for spins localized on a Bravais
lattice is defined as
\begin{equation} \label{eq:genericstructure}
  S^{\alpha\beta}(\vec{q}, \omega) = 
  \int_{-\infty}^{\infty} \!\!dt\ e^{-i\omega t} 
  \langle \mathcal S^{\alpha}(\vec{q},\!t) \mathcal S^{\beta}(-\vec{q},\!0) 
\rangle
\end{equation}
where
\begin{equation}\label{eq:spinfourier}
\mathcal S^{\alpha}(\vec{q},\!t) = 
\sum_{\vec{R}} e^{-i \vec{q}\vec{R}} \mathcal S^{\alpha}(\vec{R},\!t)
\end{equation}
is the Fourier transformation of the spin operators at lattice sites
$\vec{R}$. The superscripts $\alpha, \beta$ denote the spin
components and the brakets $\langle\cdots\rangle$ thermal
expectation values (which for $T\!=\!0$ reduce to groundstate
expectation values $\langle 0 | \cdots |0\rangle$).  Apart from known 
prefactors, Eq. (\ref{eq:genericstructure}) reflects the spectral weight 
from the magnetic neutron scattering cross section \cite{Squ78}.

If we consider transitions from the ground state $\ket{0}$ 
to some well-defined eigenstate $\ket{n}$ with energy $\omega_n(\vec{q})$, 
we obtain $\delta$-peaked contributions to the dynamical structure factor
\begin{eqnarray}
S^{\alpha\beta}(\vec{q}, \omega) &=& 
\sum_{n} \bra 0 \mathcal S^{\alpha}(\vec{q}) \ket n 
         \bra n \mathcal S^{\beta}(-\vec{q}) \ket 0\,
         \delta(\omega-\omega_n(\vec{q}))\label{eq:oned}\\
&=& \sum_{n} 
I^{\alpha\beta}_n(\vec{q})\,\delta(\omega-\omega_n(\vec{q})).
\end{eqnarray}
In interacting dimer materials, 
INS probes directly the transition from the (singlet) ground state 
$\ket{0}$ to
the lowest (triplet) excitation $\ket{t}$ 
and we will reduce our discussion to this contribution to
the dynamical structure factor. Owing to the rotational symmetry of 
the underlying Heisenberg model it is sufficient to calculate 
$I^{zz}_t(\vec{q})$ only and we use the shorthand 
$I_{sm}(\vec{q}):= I^{zz}_{t}(\vec{q})$ to denote the lowest triplet
(single magnon) contribution to the spectral weight.

The INS investigation of the materials KCuCl$_{3}$ and 
TlCuCl$_{3}$ at finite magnetic fields provides direct verification of 
this point as reported in \cite{Cavadini02/1}.

The discussion of our results is organized as follows. In section II we will
give theoretical results for a 1D array of interacting dimers. This model was
treated before \cite{Barnes99}, it is, however, instructive to demonstrate for
the simple 1D case, that existing standard expansions are modified by
additional terms which emerge starting in second order. In addition we present
the quantitative changes for the transition strength comparing first order to
10$^{\rm th}$ order results to show the effect of high order calculations. In
section III we discuss interacting dimers in a 3D network by presenting in
parallel neutron scattering results for the material $\rm KCuCl_3$ and series
expansions to 2$^{\rm nd}$ order. The same type of additional terms as in 1D
is obtained in this calculation and corrects the results for the dynamical
structure factor as obtained in the random phase approximation (RPA) before,
see Ref.~\cite{Suzuki00}. These RPA results are found to be correct only to
first order. Section IV gives our conclusions.


\section{Alternating chain}\label{sec:1d}
First we consider the one dimensional (1D) alternating $S=1/2$ spin
chain with isotropic nearest-neigbour interactions. The hamiltonian of
this model is of the following form
\begin{equation}
\label{eq:altchain}
\mathcal H = J \sum_{n=1}^{N} \left( {\mathbf S}_1(n) \cdot {\mathbf S}_2(n)
             + \lambda {\mathbf S}_2(n) \cdot {\mathbf S}_1(n+1) \right),
             \qquad J>0.
\end{equation}
Here, the alternating chain is described as a system with $N$ unit cells with
two spins each and periodic boundary conditions are used.  There are two
exchange constants, $J$ and $\lambda J$, for $\lambda=0$ the ground state of
the system consists of singlets on the intracell bonds $(n,1) - (n,2)$. These
local singlets can be excited to triplets which remain gapped excitations when
switching on $\lambda$, $0<\lambda<1$.  In the limit $\lambda=1$ we arrive at
the well known Heisenberg antiferromagnet (HAFM) with pairs of $S=1/2$ spinons
as lowest gapless excitations. Other related models are described in
Ref.~\cite{Brehmer96}.

The triplet excitation energies $\omega(q)$ have been obtained by perturbation
expansion in $\lambda$ up to $9^{\rm th}$ order in Ref.~\cite{Barnes99} and to
$10^{\rm th}$ order in Ref.~\cite{MMueller00} using the cluster expansion
approach.

\subsection{The dynamical structure factor}
Turning to the calculation of the structure factor for a system with
hamiltonian \eref{eq:altchain}, we note 
(see Eqs.~(\ref{eq:genericstructure},\ref{eq:spinfourier}))
that for this calculation we have
to specify the positions of the spins, ${\vec R}$, in space (whereas the
eigenvalues depend solely on the exchange constants). In slight generalization
of a strictly linear geometry we allow for our calculation the separation
${\vec d}$ of two spins in one unit cell to be different in magnitude and
direction from the separation ${\vec a}$ of two adjacent spins in different
unit cells (note that ${\vec a}$ defines the overall chain direction). The
resulting geometry is shown in \Fref{fig:dimerstruct} and makes clear the
relation to the real 3D systems to be dealt with in the next section: the 1D
chain defined in Eq. (\ref{eq:altchain}) can alternatively be looked at as a
ladder with rung and diagonal interactions only. For the chain geometry shown
in \Fref{fig:dimerstruct} the spectral intensity up to first order in
$\lambda$ is obtained as follows
\begin{equation}
I_{sm}(\vec{q}) \:=\: \sin^2\!\frac{\vec{q}\vec{d}} 2 
   \big(1+\frac 1 2 \lambda\cos(\vec{q}\vec{a})\big) +\mathcal O(\lambda^2).
\end{equation}
Here, $\vec{q}$ is the wave vector, $\vec d$ and $\vec a$ the separation 
of the spin sites within and between the dimers, respectively. 
We note that the term $\propto \sin^2(\vec{q}\vec{d}/2)$ is
typical for systems consisting of isolated dimers, it is known as the
dimer structure factor \cite{Furrer77}. The first order correction in 
$\lambda$ adds an additional modulation to the intensity, which depends 
on the ratio $\sigma=\|\vec{d}\|/\|\vec{a}\|$.

Using the cluster expansion method (see \ref{clexpansion}) we
have systematically calculated the series in $\lambda$ for the
intensity up to the tenth order. This requires linked clusters
consisting of maximum 10 bonds. The resulting series can be split into
three different terms:
\begin{equation}\label{eq:1dform}
I_{sm}(\vec{q}) \:=\: B_c(\vec{q},\lambda)+B_s(\vec{q},\lambda) 
            + \Lambda(\vec{q},\lambda).
\end{equation} 
To illustrate the result we give the series up to fourth
order\footnote{Higher order terms are available on request.}:
\begin{equation}
\eqalign{
B_c(\vec{q},\lambda) &=\sin^2\!\frac{\vec{q}\vec{d}} 
2\sum_{j=0}^4\mu_j\cos(j\vec{q}\vec{a}),\\
B_s(\vec{q},\lambda) &=\sin\vec{q}\vec{d}\sum_{j=0}^4\nu_j\sin(j\vec{q}\vec{a})
}
\end{equation}
where
\begin{eqnarray}
\ms \mu_0  = 1 - \frac 5{16}\lambda^2 - \frac 3 {32} \lambda^3 
+\frac{25}{1536}\lambda^4,\quad\nu_0  = 0, \label{eq:mu0}\\
\ms \mu_1 = \frac 1 2 \lambda - \frac 1 8\lambda^2 - \frac 
5{192}\lambda^3 + \frac{41}{2304}\lambda^4,
\quad\nu_1 = \frac 1 8 \lambda^2 +\frac 7 {192}\lambda^3 
- \frac{131}{4608}\lambda^4,\nonumber\\
\ms \mu_2  = \frac 3 {16}\lambda^2 + \frac 7 {48}\lambda^3 + 
\frac{23}{1024}\lambda^4, \quad\nu_2  = \frac 1 {96}\lambda^3 
+ \frac{25}{4608}\lambda^4,\nonumber\\
\ms \mu_3  = \frac 5 {64}\lambda^3 + \frac{155}{2304}\lambda^4,
\quad \nu_3  = \frac{23}{2304}\lambda^4,\nonumber\\
\ms \mu_4=\frac{35}{1024}\lambda^4,\quad\nu_4=0\nonumber,
\end{eqnarray}
and
\begin{equation}
\Lambda(\vec{q},\lambda) = \frac 1 {128}\lambda^4\big(\cos(2\vec{q}\vec{d})
-\cos(2\vec{q}\vec{a})\big).
\end{equation}
The terms in $B_c(\vec{q},\lambda)$ are consistent with previous
publications \cite{Barnes99}, whereas $B_s(\vec{q},\lambda)$ and
$\Lambda(\vec{q},\lambda)$ contain additional corrections. They originate
from a complete expansion of both the ground state $\ket{0}$ and the
first excited state $\ket{t}$. If one assumes that
$\bra{0}S_{2i}\ket{t}=-\bra{0}S_{2i+1}\ket{t}$ for the matrix elements
on even and odd sites, one ends up with $B_c(\vec{q},\lambda)$
only. However this is only correct to first order and in general we
have $\bra{0}S_{2i}\ket{t}\not=-\bra{0}S_{2i+1}\ket{t}$. This
inequality arises from virtual states with odd parity under exchange 
of two triplets which occur during the perturbation expansion for the 
first time in second order. 

\begin{figure}
\begin{center}
    \includegraphics[width=14cm]{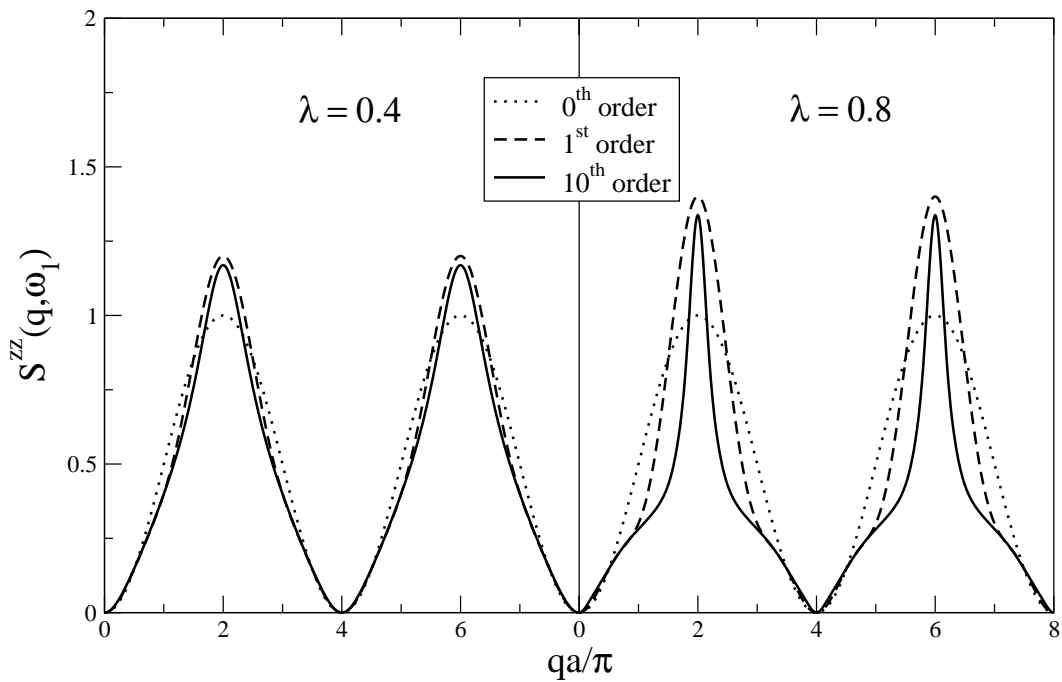}
    \caption{\label{fig:sf1020}Structure factor with
    $\lambda=0.4,\,0.8$, $\vec a \:\|\: \vec d$ and ratio $d/a=10/20$.}
\end{center}
\end{figure}

\begin{figure}
\begin{center}
    \includegraphics[width=14cm]{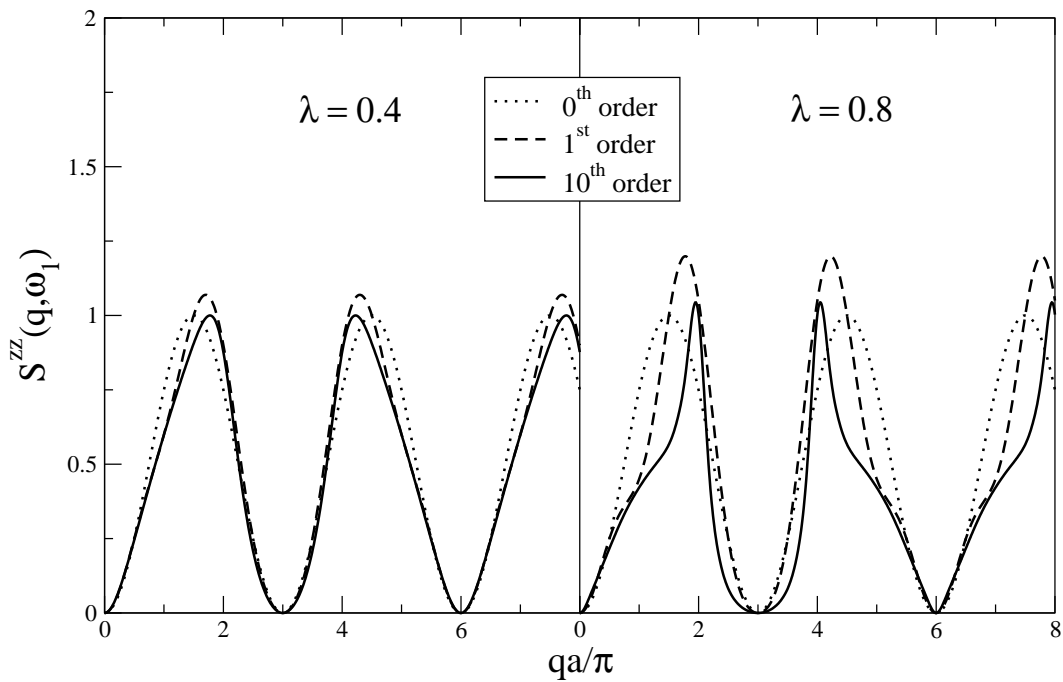}
    \caption{\label{fig:sf1015}Structure factor with 
$\lambda=0.4,\,0.8$, $\vec a \:\|\: \vec d$ and ratio $d/a=10/15$.}
\end{center}
\end{figure}

In \Fref{fig:sf1020} and \Fref{fig:sf1015} we show some typical plots of the
intensity $I_{sm}(\vec{q})$ for two different ratios $\sigma = 10/15,\,10/20$
and two different coupling strength $\lambda = 0.4,\,0.8$ and strictly linear
geometry, $\vec d \| \vec a$ (then, only the component of wavevector in chain
direction enters). The difference between the zeroth and the first
order emerge very clearly. The higher order terms emphasize the modulation
originating from the two length scales $\|\vec{a}\|$ and $\|\vec{d}\|$.
\subsection{Sum rule}
The total integrated scattering intensity has a well defined magnitude,
determined by the local spin length through the following sum rule:
\begin{equation}\label{eq:sumoned}
\mathcal I = 
\sum_{\alpha}\frac{\int\!d\vec{q}\int\!\frac{d\omega}{2\pi}
       S^{\alpha\alpha}(\vec{q},\omega)}{\int\!d\vec{q}}=S(S+1). 
\end{equation}
For the one dimensional alternating chain the contribution from the one-magnon
part to the total spectral weight is calculated from eq.~\eref{eq:1dform};
the integral reduces to the constant part $\mu_0$ of \eref{eq:mu0}, since only
non oscillating terms survive the integration:
\begin{equation}\label{eq:sumrule}
{\mathcal I}_{sm} = \frac 3 4 \left( 1 - \frac 5{16}\lambda^2  
      - \frac 3 {32} \lambda^3 +\frac{25}{1536}\lambda^4  
      + \ldots \right) \le \frac 3 4 
      \qquad S=\frac 1 2,\quad \lambda\ll 1.
\end{equation}
For the noninteracting case ($\lambda=0$) the sum rule is exhausted
by the one triplet excitation since it is an exact eigenstate: 
From \eref{eq:oned} we see that this excitation gives the only 
non-vanishing 
matrix element. Switching on the coupling between the dimers, more 
and more intensity goes in two or more magnon scattering processes.
A theoretical discussion of the multimagnon states for the one dimensional 
alternating chain is given in Refs.~\cite{Barnes99,Damle98,Trebst00}.

                
\section{The 3D dimer substances KCuCl$\mathbf{_3}$ and TlCuCl$\mathbf{_3}$}
In this section we extend the calculation of singlet-triplet
intensities to three dimensional substances such as KCuCl$_3$ and
TlCuCl$_3$. These compounds are weakly interacting quantum spin
systems which exhibit an excitation gap. Similar to the alternating 
chain
discussed in the previous section this is based on the existence of
strongly interacting bonds forming dimers. In these materials the
orientation af the dimers alternates, i.\,e. each unit cell consists
of four spins with two differently oriented dimers.

The following considerations are valid for dimer systems
composed of two dimers per unit cell. We write for the Fourier
transformed spin operators
\begin{equation}\label{eq:spinfourier3d}
\fl S^{z}\!(\vec{q}) = \frac 1 {\sqrt{2N}}\sum_{\vec{n}}\sum_{k=1}^2 
    e^{-i\vec{q}(\vec{n}+\vec{R}_k)}\!\!
    \left[e^{-i\frac{\vec{q}\vec{d}_k} 2} 
    S^z\!(\vec{n}\!+\!\vec{R}_k\!+\!\frac{\vec{d}_k} 2) 
    +e^{i\frac{\vec{q}\vec{d}_k} 2}  
    S^z\!(\vec{n}\!+\!\vec{R}_k\!-\!\frac{\vec{d}_k} 2) \! \right],
\end{equation}
where $\vec{n}$ denotes the unit cell, $\vec{R}_k$ the center of the
dimer and $\vec{d}_k$ the separation of the two spins forming the
dimer. The first sum in Eq.~\eref{eq:spinfourier3d} is taken over all
unit cells. In the case of weakly interacting dimers the localized
triplet states are replaced by Bloch-like triplet modes which
propagate due to the interaction network between the dimer units. 
The details of the explicit
calculation of the transition matrix elements for a Bravais dimer lattice 
are illustrated in Ref.~\cite{Barnes99}.

In Tab.~\ref{tab:ww} we describe the interaction network listing the basic
lattice vectors associated with nonzero interdimer interactions for the
materials considered. In Tab.~\ref{tab:wwvalues} we give the numerical values
of the intradimer exchange $J$ (in meV) and of the interdimer exchange
interactions (in relative units) for the compounds KCuCl$_3$ and TlCuCl$_3$. 
(Slightly improved values for KCuCl$_3$ have been determined in \cite{MMuellerDiss02}, the 
difference, however, is not visible in the Figures \ref{fig:theory} and \ref{fig:fit} below.)
\begin{table}
\caption{\label{tab:ww}Considered interactions in KCuCl$_3$
(and TlCuCl$_3$). Vectors $\vec{g}_k$ denote the distances between 
the dimer centers.} 
\begin{indented}
\item[]\begin{tabular}{@{}lcc}
\br
distance $\vec{g}_k$ &\centre{2}{Interactions between}\\
between dimers&equivalent spins ($i=j$)&non equivalent spins ($i\not= j$)\\
\mr
$\vec{g}_1= \vec{a}$           &$J_{(100)}$&$J^{\prime}_{(100)} $ \\
$\vec{g}_2=2\vec{a}+\vec c$ &$-$&$J^{\prime}_{(201)} $ \\
$\vec{g}_3= \vec{a}+\frac 1 2 (\vec{b}+\vec{c})$ &$J_{(1\frac 1 
2\frac 1 2)}$&$J^{\prime}_{(1\frac 1 2\frac 1 2)}$ \\
$\vec{g}_4= \vec{a}-\frac 1 2 (\vec{b}-\vec{c})$ &$J_{(1\frac 1 
2\frac 1 2)}$&$J^{\prime}_{(1\frac 1 2\frac 1 2)}$ \\
\br
\end{tabular}
\end{indented}
\end{table}

\begin{table}
\caption{\label{tab:wwvalues}Values of exchange interactions in KCuCl$_3$\cite{MMueller00}
and TlCuCl$_3$\cite{MMueller02}. The intradimer interaction $J$ is given in meV, interdimer 
interactions are given in units of $J$ for the respective compound}
\begin{indented}
\item[]\begin{tabular}{@{}lcc}
\br
exchange constant& KCuCl$_3$ & TlCuCl$_3$  \\
\mr
$J$         &  4.25 meV & 5.68 meV \\
$J_{(100)}$ & 0.00 & 0.06 \\
$J^{\prime}_{(100)}$ & 0.10 & 0.30\\
$J^{\prime}_{(201)} $ & 0.18 & 0.45 \\
$J_{(1\frac 1 2\frac 1 2)}$ & 0.20 & 0.16 \\
$J^{\prime}_{(1\frac 1 2\frac 1 2)}$ & 0.05 & -0.10 \\
\br
\end{tabular}
\end{indented}
\end{table}

The different dimer orientations in the unit cell do not affect the dimer
lattice directly, i.\,e. the lowest excitation does not depend on the dimer
orientation. But the full translational symmetry is obviously reduced if the
dimer sites are distinct by their orientation.

In analogy to phonons in a lattice with a basis there will be two 
excitation modes. Therefore we use the following zeroth order ansatz 
for the one triplet wave function, which manifest translational symmetry:
\begin{equation}
\ket{\vec{q}}^{(0)} = \frac 1 {\sqrt{2N}} \sum_{k=1}^2\sum_{\vec{n}} 
     c_k e^{-i\vec{q}(\vec{n}+\vec{R}_k)}\ket{\vec{n} + 
\vec{R}_k}^{(0)}.
     \label{eq:q0}
\end{equation}
The states $\ket{\vec{n} + \vec{R}_k}^{(0)}$ denote a triplet at site
$\vec{n} + \vec{R}_k$ with all the other sites occupied by
singlets. We have introduced the coefficient $c_k$ to take into
account the different dimer orientations. The $c_k$'s are determined
requiring that $c_1\ket{\vec{n} + \vec{R}_1}+c_2\ket{\vec{n} +
\vec{R}_2}$ is diagonal in the subspace of the one triplet
excitations. Two solutions ($\pm$) are obtained:
\begin{equation}
(c^+_1,c^+_2) = (1,1) \qquad {\rm or} \qquad(c^-_1,c^-_2) = (1,-1).
\end{equation}
The solution $c^-$ is connected to umklapp scattering where 
$\vec{q} \longrightarrow \vec{q} + \vec{u}$. Umklapp processes are possible in
KCuCl$_3$ and TlCuCl$_3$ with an integer number of the reciprocal
lattice vector $\vec{u}=\vec{b}^{\ast}$ or $\vec{u}=\vec{c}^{\ast}$.

The resulting energies are denoted by $\omega^{\pm}(\vec{q})$ where 
we describe both modes in the first cristallographic Brillouin zone. 
Up to first order we obtain the well known dispersion relation 
\cite{MMueller00,Cavadini00/1} in units of $J$:
\begin{equation}\label{eq:omega}
\omega^{\pm}(\vec{q}) = 1 + 2\sum_{i=1}^2\beta_i\cos(\vec{g}_i\vec{q}) 
      \pm 2\sum_{i=3}^4\beta_i\cos(\vec{g}_i\vec{q}) + \mathcal 
O(\lambda^2).
\end{equation}
Here and in the following we use some short-hands of various
combinations of coupling constants ($\epsilon_i$ and $\gamma_i^j$ 
will be needed for the intensity calculation below):
\begin{eqnarray}
\fl\beta_1 =  \frac 1 4(2J_{(100)}-J'_{(100)}),&\quad\varepsilon_1 
    =\frac 1 4(2J_{(100)}+J'_{(100)}),&\quad\gamma_1 = \frac 1 4 
J'_{(100)},\nonumber\\
\fl\beta_2 = -\frac 1 4 J'_{(201)},&\quad\varepsilon_2 
    =\frac 1 4 J'_{(201)},&\quad\gamma_2 = -\frac 1 4 J'_{(201)},\\
\fl\beta_3 =  \frac 1 4(J_{(1\frac 1 2\frac 1 2)}-J'_{(1\frac 1 2\frac 
1 2)}),&\quad \varepsilon_3 = \frac 1 4(J_{(1\frac 1 2\frac 1 2)} 
        +J'_{(1\frac 1 2\frac 1 2)}),&\quad\gamma^{\pm}_3 = 
        \frac 1 4(\pm J_{(1\frac 1 2\frac 1 2)} 
        + J'_{(1\frac 1 2\frac 1 2)}),   \nonumber\\
\fl\beta_4 =\beta_3,&\quad\varepsilon_4 =\varepsilon_3
   ,&\quad\gamma^{\pm}_4 =\gamma^{\pm}_3.  \nonumber
\end{eqnarray}
To obtain the transition matrix element one has to expand both the
ground state and the one triplet state perturbatively in the coupling
constants. To indicate the order we take all interdimer couplings 
proportional to a constant $\lambda$. 

%
%

\subsection{The ground state}
There are four different directions $\vec{g}_i$ in which we find interdimer 
interactions (see Tab.~\ref{tab:ww}). As well as in the one dimensional case the
unperturbed ground state is a product of singlets placed on the rungs:
\begin{equation}\label{eq:gznull}
\ket{G}^{\!(0)}
=\prod_{\vec{n}}\ket{s_{\vec{n}+\vec{R}_1}}\ket{s_{\vec{n}+\vec{R}_2}}=\ket{S}.
\end{equation}
Due to the different orientations we distinguish between singlets at
$\vec{n}+\vec{R}_1$ and $\vec{n}+\vec{R}_2$. Up to the second order
the ground state including all relevant states for the
structure factor is
\begin{eqnarray}\label{eq:gzzwei}
\ket{G}^{(2)} =& \alpha_0\ket S  
  +\frac{\sqrt 3} 2 \sum_{i=1}^4 \sum_{k=1}^2\sum_{\vec{n}} 
   \beta_i(1+\varepsilon_i)\ket{\vec{n}\!+\!\vec{R}_k,\vec{n} 
   \!+\!\vec{R}_k\!+\!\vec{g}_i}^{(0,0)}\\
  &-\frac{\sqrt 3} 2 \sum_{i,j=1}^4 \sum_{k=1}^2 \sum_{\vec{n}}
   \beta_i\beta_j\ket{\vec{n}\!+\!\vec{R}_k,\vec{n} 
   \!+\!\vec{R}_k\!+\!\vec{g}_i\!+\!\vec{g}_j}^{(0,0)}\\
  &-\frac{\sqrt 3} 2 \sum_{{i,j=1}\atop{i\not=j}}^4  
   \sum_{k=1}^2 \sum_{\vec{n}}\beta_i\beta_j\ket{\vec{n}
\!+\!\vec{R}_k,\vec{n}\!+\!\vec{R}_k\!+\!\vec{g}_i\!-\!\vec{g}_j}^{(0,0)}\\
&+\ {\rm states\ with\ three\ or\ four\ triplet\ excitations}.
\end{eqnarray}
The indices $i,j$ are linked to the different interaction directions and $k$ 
counts the two dimer sites in the elementary cell. We further denote the 
states having two triplets at sites $\vec{r}$ and $\vec{r'}$ with well-defined 
$S_{\rm tot}$ and $S^z_{\rm tot}$ by
$\ket{\vec{r},\vec{r'}}^{(S_{\rm tot},S^z_{\rm tot})}$. $\alpha_0$ is 
a normalization factor which guarantees that $\langle G\ket{G}^{(2)} 
= 1 + \mathcal O(\lambda^3)$:
\begin{equation}
\alpha_0 = 1 - \frac 3 4 N \sum_{i=1}^4 \beta^2_i.
\end{equation} 
Note that $N$ labels the number of unit cells, whereas $2N$ is the
number of dimers in the system.

%
%

\subsection{One triplet excitation}
The expansion of the wave function for the one triplet excitation in
the interdimer interactions is obtained to first order 
by acting with the Hamiltonian $\mathcal H_1$ on the state
\eref{eq:q0} and some subsequent normalization. 
In addition to simple propagation of the triplets this
leads to the generation of two triplet excitations
$\ket{\vec{r},\vec{r'}}^{(1,0)}$ where $(S_{\rm tot},S^z_{\rm tot}) =
(1,0)$ is the total spin and total magnetization and
$\vec{r},\vec{r'}$ label the lattice sites occupied by triplets. 
There are also three triplet excitations with
the same spin quantum numbers. Second order terms contribute to the
third order\footnote{All contributing terms can be obtained on
request.} of the intensity only and will not be calculated
here. However, normalization of the wave function has to be done up to
second order terms.


\subsection{The dynamical structure factor}
To leading order one expects a dimer-like structure factor as in
Sec.~\ref{sec:1d}. In fact there are two different contribution due to
the two dimer structure in the elementary cell:
\begin{equation}\label{eq:ifirstorder}
I_{\pm}(\vec{q})  = D_{\pm}^2(\vec{q}) + \mathcal O(\lambda^1) =   
\left[
\sin\frac{\vec{q}\vec{d}_1} 2 \pm \sin\frac{\vec{q}\vec{d}_2} 
2\right]^2 
+ \mathcal O(\lambda^1).
\end{equation}
The indices $\pm$ refer to \eref{eq:omega} and correspond to symmetric resp.
antisymmetric modes for $\vec{q}$ in the first Brillouin zone. However, the
role of symmetric and antisymmetric modes is interchanged for $\vec{q} 
\longrightarrow \vec{q} +\vec{\tau}$ where 
$e^{i\vec{\tau}(\vec{R}_2-\vec{R}_1)} = -1$. From now on $\vec{q}$ stays in the
first Brillouin zone.

The representation of 
\eref{eq:ifirstorder} is instructive in order to emphasize that in general
there are contributions from two different modes for a wavevector $\vec{q}$
in the crystallographic Brillouin zone. The total contribution 
\begin{equation}
I_+(\vec{q})+I_-(\vec{q})=2\sin^2\frac{\vec{q}\vec{d}_1} 2
+2\sin^2\frac{\vec{q}\vec{d}_2} 2
\end{equation}
reproduces correctly the structure factor $\int d\omega S(\vec{q},\omega)$ as
calculated from \eref{eq:genericstructure} in lowest order.

A finite contribution for the excitation mode with energy 
$\omega^-(\vec{q})$ requires $\vec{q}\vec{d}_1 \neq \vec{q}\vec{d}_2$. 
Taking into account the dimer orientations $\vec{d}_{1,2}$ where 
\begin{eqnarray}
\vec{d}_1=0.48\vec{a}+0.10\vec{b}+0.32\vec{c}, \nonumber\\  
\vec{d}_2=0.48\vec{a}-0.10\vec{b}+0.32\vec{c}
\end{eqnarray}
we deduce the condition $\vec{q}_b \neq 0$, in agreement with the experimental
observations.

Considering higher order corrections to the ground state and the first
excited state as presented below, 
we get the following result valid up to second order:
\begin{eqnarray}
I_{\pm}(\vec{q}) &=&  \frac 1 4 D_{\pm}^2(\vec{q})
    \big(1 - \Omega_{\pm}(\vec{q}) + \Omega^2_{\pm}(\vec{q}) 
    -\Sigma_{\pm}(\vec{q})-\Sigma_1(\vec{q})\big) \nonumber\\
  &+&\frac 1 2 D_{\pm}(\vec{q})
    \big(\cos\frac{\vec{q}\vec{d}_1} 2  \pm \cos\frac{\vec{q}\vec{d}_2} 2 \big) \Delta_{\pm}(\vec{q})
\end{eqnarray}
where
\begin{eqnarray}\label{eq:strukturresult2}
\fl\Omega_{\pm}(\vec{q})   = 2\sum_{i=1}^2 \beta_i\cos(\vec{q}\vec{g}_i)\pm2\sum_{i=3}^4\beta_i\cos(\vec{q}\vec{g}_i),\label{eq:mean}\\
\fl\Sigma_{\pm}(\vec{q}) = \sum_{i=1}^4\left[3\beta_i^2 
   -\beta_i^2\cos(2\vec{q}\vec{g}_i)\right] + 2\sum_{i=1}^2 \beta_i\varepsilon_i\cos(\vec{q}\vec{g}_i)\pm2\sum_{i=3}^4\beta_i\varepsilon_i\cos(\vec{q}\vec{g}_i),\\
\fl\Sigma_1(\vec{q}) = 4\sum_{i=1}^2\gamma_i^2 
   \left[1+\cos(\vec{q}\vec{g}_i)\right]
   +2\sum_{i=3}^4\left[{\gamma_i^+}^2+{\gamma_i^-}^2 
      + 2\gamma_i^+\gamma_i^-\cos(\vec{q}\vec{g}_i)\right],\\
\fl\Delta_{\pm}(\vec{q}) = 2\sum_{i=1}^2\beta_i\gamma_i\sin(\vec{q}\vec{g}_i)
   +2\sum_{i=3}^4\beta_i(\gamma^-_i\pm\gamma^+_i)\sin(\vec{q}\vec{g}_i)
\end{eqnarray}
The dynamical structure factor contains contributions from both
excitation modes:
\begin{equation}\label{eq:strures}
S^{zz}(\vec{q},\omega) = 
   I_{+}(\vec{q}) \: \delta\!\left(\omega-\omega^+(\vec{q})\right) 
   + I_{-}(\vec{q}) \: 
\delta\!\left(\omega-\omega^-(\vec{q})\right),
\end{equation}
where $\omega^{\pm}(\vec{q})$ denotes the one triplet excitation 
energy as in \eref{eq:omega}.
We note that $\Omega(\vec{q})$ is the energy of the one triplet 
excitation $\omega^+(\vec{q})$ up to first order. If we neglect 
the terms $\Sigma_i(\vec{q})$ and $\Delta_{\pm}(\vec{q})$ the 
result reduces to RPA-like calculations \cite{Suzuki00} where $
I_{\pm}(\vec{q}) \propto 1 / \omega^{\pm}(\vec{q})$. 

\begin{figure}
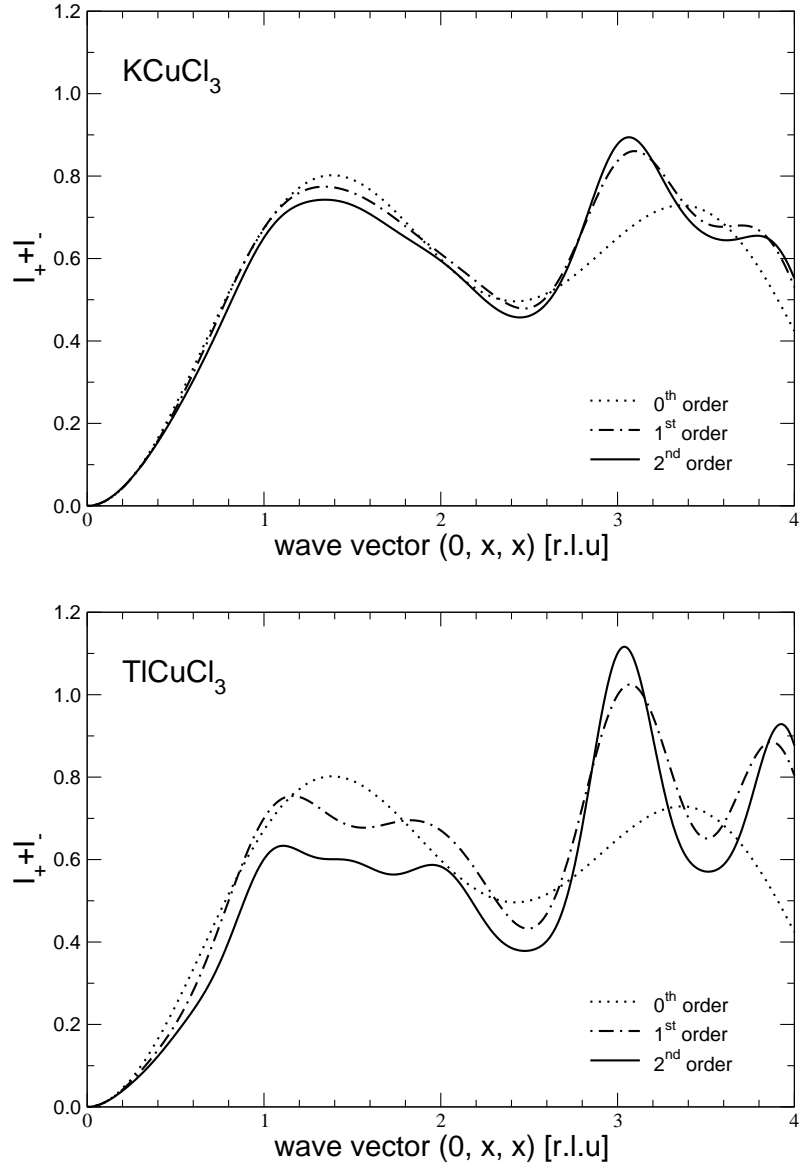

\begin{center}
    \includegraphics[width=10.5cm]{Figure4c_rev1.eps}\\[5mm]
    \includegraphics[width=10.5cm]{Figure4d_rev1.eps}
\caption{\label{fig:theory}Total spectral weight $I_+ + I_-$ along the 
$(0,x,x)$ direction in 0$^{th}$, 1$^{st}$ and 2$^{nd}$ order.
Top panel: KCuCl$_{3}$, bottom panel: TlCuCl$_{3}$.}
\end{center}
\end{figure}

In order to demonstrate the effect of interdimer interactions on the
dynamical structure factor we show in \Fref{fig:theory} theoretical
results for KCuCl$_{3}$ (top panel) and TlCuCl$_{3}$ (bottom panel) in
0$^{\rm th}$ (noninteracting dimers), 1$^{\rm st}$ and 2$^{\rm nd}$
order, exchange parameters are taken from Tab.~\ref{tab:wwvalues}. As
in the INS experiments to be discussed in the next subsection and as
in Ref.~\cite{Suzuki00}, the variation of the spectral weight of the
triplet excitation $I_+ + I_-$ with wave vector is shown along the
$(0,x,x)$ direction of reciprocal space, such that both modes
contribute with finite weight. Evidently, higher order corrections are
more important for TlCuCl$_{3}$ with its larger exchange constants,
but even for these larger values the comparison of different orders seems to
indicate convergence. Exhaustive experimental results for both $I_+$ and 
$I_-$ are available for KCuCl$_{3}$ and will be discussed in the next 
subsection. Clearly, devoted INS experiments for TlCuCl$_{3}$ are of 
considerable interest (see also \cite{Cavadini00/1} and \cite{MMueller02}); 
the ratio of spectral weights at wave vectors corresponding
to maximum and minimum intensity appears to be a reasonable quantity
to test the agreement with our theoretical results.  Present results
do not indicate that higher than 2$^{\rm nd}$ order calculations are
required.

\subsection{Experiment}
The INS results on the material KCuCl$_{3}$ were collected at the IN3 neutron
spectrometer, Institut Laue-Langevin, Grenoble (France).  Standard focussing
geometry was adopted for all energy scans performed under constant final
energy $E_{f}=13.7$meV. A pyrolitic graphite (PG) filter in front of the
analyser was further used to suppress higher order contaminations.

The INS profiles were obtained at fixed $T=2$K for the wave vectors along the
$(0,x,x)$ direction of reciprocal space, which is suited to demonstrate the
issues introduced above.  For this purpose, a KCuCl$_{3}$ single crystal was
aligned for scattering in the $b^{*}c^{*}$ plane.  The spectral weight of the
triplet excitation was determined from global least squares fits to the
measured neutron profiles, assuming gaussian peaks on top of a common
background. The center of the peaks was further fixed at the energies
$\omega^{\pm}(\vec{q})$ resulting from the analysis presented in
Ref.~\cite{MMueller00}. (see Tab.~\ref{tab:wwvalues})
\begin{figure*}
    \includegraphics[width=14cm]{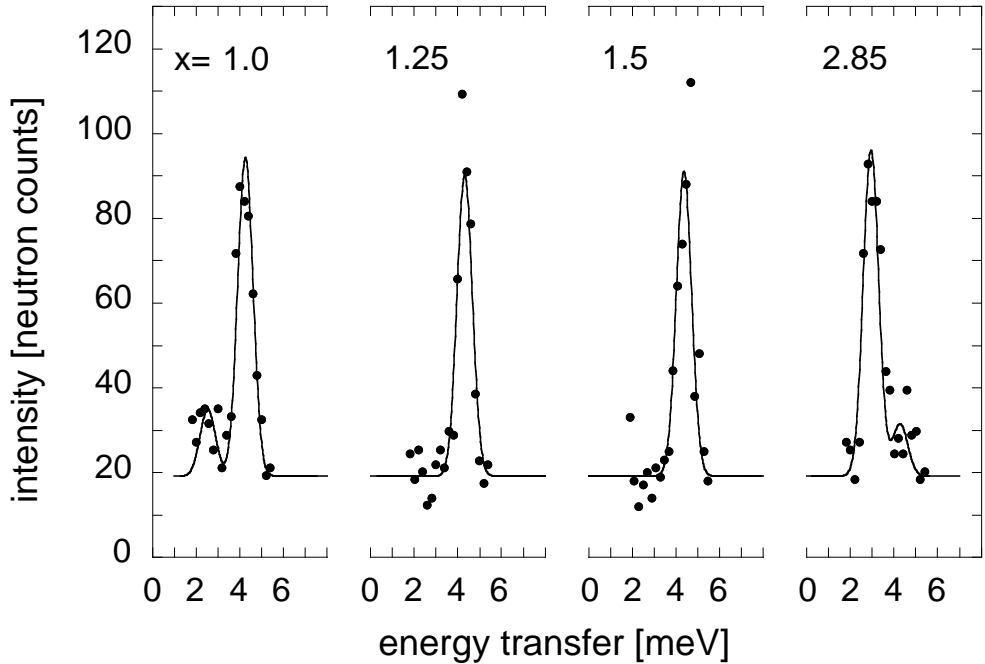}
    \caption{\label{fig:profile}Typical fits of the neutron profiles 
    for wave vectors $\vec{q}$ taken at selected $(0,x,x)$ values in (r.l.u)}
\end{figure*}
Our present results complete the experimental investigation of the
$b^{*}c^{*}$ plane summarized in Ref.~\cite{Cavadini00/1}, and references
therein.

In \Fref{fig:profile} typical fits of the neutron profiles are shown for wave
vectors $\vec{q}$ at selected $(0,x,x)$ values, in reciprocal lattice units
(r.l.u.) of the unit cell. In accordance with the theoretical expectations,
both excitation modes $\omega^{\pm}(\vec{q})$ are visible along $(0,x,x)$ but
the spectral weight $I_{\pm}(\vec{q})$ strongly depends on $x$.  Continuous
lines denote the global least squares fit function, symbols the profiles in
neutron counts. The statistical tolerance scales according to the neutron
counts.
\begin{figure}
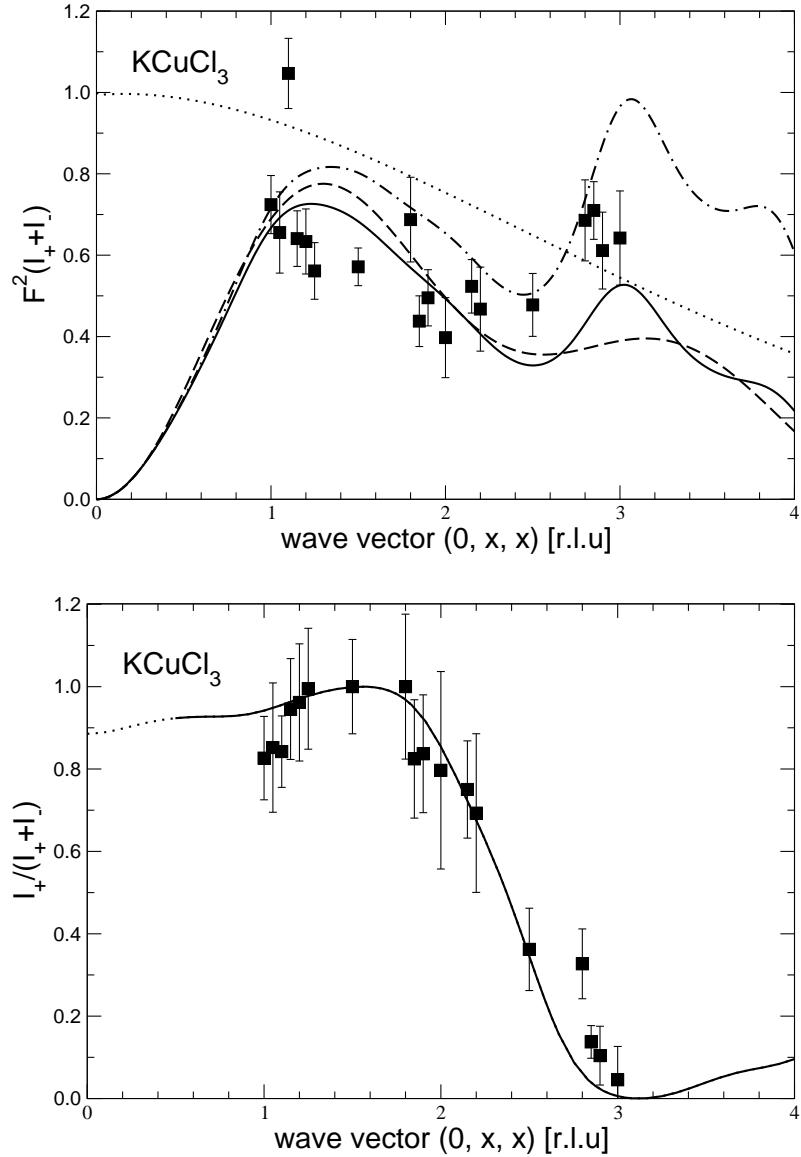

\begin{center}
    \includegraphics[width=10.5cm]{Figure4a_rev1.eps}\\[5mm]
    \includegraphics[width=10.5cm]{Figure4b_rev1.eps}
\caption{\label{fig:fit}Spectral weight of the triplet excitation for KCuCl$_3$
  along the $(0,x,x)$ direction in reciprocal space.
  Symbols indicate experimental data from least squares fit to the 
  profiles. Top panel: theoretical result for $F^2 (I_+ + I_-)$ in second order
  (full line) and for noninteracting dimers (dashed line), $F^2$ (dotted line)
  and $(I_+ + I_-)$ in second order (dash-dotted line).
  Bottom panel, full line: 
  theoretical result in second order for the relative spectral weight.}
\end{center}
\end{figure}
In \Fref{fig:fit}, the fitted spectral weight is compared to the model
expectations previously introduced. In the top panel, the total
spectral weight $F^{2}(I_{+}+I_{-})$ (full line) is compared to the
experimental observations (symbols). The only free parameter is an
overall scaling factor accounting for the size of the sample, both the
plain calculation (dashed dotted line, second order) and the squared
magnetic form factor $F^{2}$ (dotted line) \cite{Brown92} are shown
separately for convenience. In the bottom panel, the relative spectral
weight $I_{+}/(I_{+}+I_{-})(\vec{q})$ is compared to the experimental
observations, as indicated. The graphical representation underlines
the redistribution of the spectral weight among $I_{+}$ and $I_{-}$
which occurs along the $(0,x,x)$ direction of reciprocal space.

From \Fref{fig:fit}, reasonable agreement between predictions from the
dimer model and experimental results is concluded. The spectral weight
is dominated by the bipartite dimer structure, which governs the
result in the noninteracting dimer limit, but the existence of higher 
order corrections is clearly seen close to the second maximum 
($x \approx 3$). Details of higher
order corrections remain almost beyond statistics for KCuCl$_3$ but
may become more pronounced in the sister material TlCuCl$_{3}$. Our
results improve on the previous RPA calculations \cite{Suzuki00}
which are correct only to first order. Related investigations along
different directions of the $b^{*}c^{*}$ plane were successfully
compared to RPA calculations in Ref.~\cite{Cavadini00/2}.  The
relative spectral weight (bottom panel of \Fref{fig:fit}) is very well
described already in the noninteracting dimer picture (not shown 
in \Fref{fig:fit}), higher order corrections are below statistical 
significance for this quantity.


\subsection{Sum rule}
We calculate the one magnon contribution to the total integrated scattering
intensity in order to check the sum rule \eref{eq:sumrule}. As seen in
\eref{eq:strures} the dynamical structure factor consists of two parts.
Integrating over $\vec{q}$, only non oscillating terms survive, giving

\begin{equation}\label{eq:sumresult}
{\mathcal I}_{sm} = \frac 3 4 \left[ 1 
     -\sum_{i=1}^4\beta_i^2 - 4 (\gamma_1^2+\gamma_2^2) 
     - 2({\gamma_3^+}^2+{\gamma_4^+}^2+{\gamma_3^-}^2+{\gamma_4^-}^2) 
\right] + \mathcal O(\lambda^3).
\end{equation}

Using the coupling constants as calculated in Ref.~\cite{MMueller00} we
estimate here the intensity which goes into higher order scattering processes
like two or more magnon scattering. We obtain ${\mathcal I}_{sm} = 0.7217$
which means that $96.23$\% of the total scattering intensity is concentrated
in the lowest triplet excitation. Although the interactions in TlCuCl$_{3}$
are more pronounced most of the scattering intensity still goes in the one
magnon process which is reflected by inserting calculated coupling constants
\cite{MMueller02} in \eref{eq:sumresult}: ${\mathcal I}_{sm} = 0.7021$ or
$93.62$\% respectively. The absolute experimental determination of the
spectral weight from dimers is exemplified in Ref.~\cite{Zheludev96}, but a
devoted investigation of the materials KCuCl$_{3}$ and TlCuCl$_{3}$ has not
been performed up to now.

\section{Conclusion}

We have presented series expansions for the dynamical structure factor valid
generally for lattices with two dimers per unit cell. Applying our results to
the interacting dimer material KCuCl$_{3}$, we have shown that results
obtained by inelastic neutron scattering are reasonably well described by the
theoretical calculations. Our expressions apply as well to the sister material
TlCuCl$_{3}$, which shares with KCuCl$_{3}$ the structure of the exchange
couplings, but has larger exchange strengths. For the specific direction in
$\vec q$-space considered here our results show that higher
order terms are not relevant for relative spectral weights (see \Fref{fig:fit}
as measured in KCuCl$_3$ with present intensity) and we expect that this is 
generally true. Second order shifts, however, show up in absolute spectral 
weights \cite{MMuellerDiss02}, most clearly in TlCuCl$_{3}$.

The materials KCuCl$_{3}$ and TlCuCl$_{3}$ recently have been demonstrated to
undergo field-induced magnetic ordering. The evolution of the excitation modes
at finite magnetic field has been described in a comprehensive theoretical
study \cite{Matsumoto02}, albeit limited to the energy of the excitations.
Theoretical investigations of the spectral weight in an external magnetic
field are now under preparation.

                
\appendix
\section{Cluster Expansion}\label{clexpansion}
In this appendix we briefly summarize the method of cluster expansion 
for the dynamical structure factor in the case of the alternating 
chain. Some detailed considerations can be found 
elsewhere \cite{Gelfand00}. 
\begin{figure}
\begin{center}
\includegraphics[width=7.5cm]{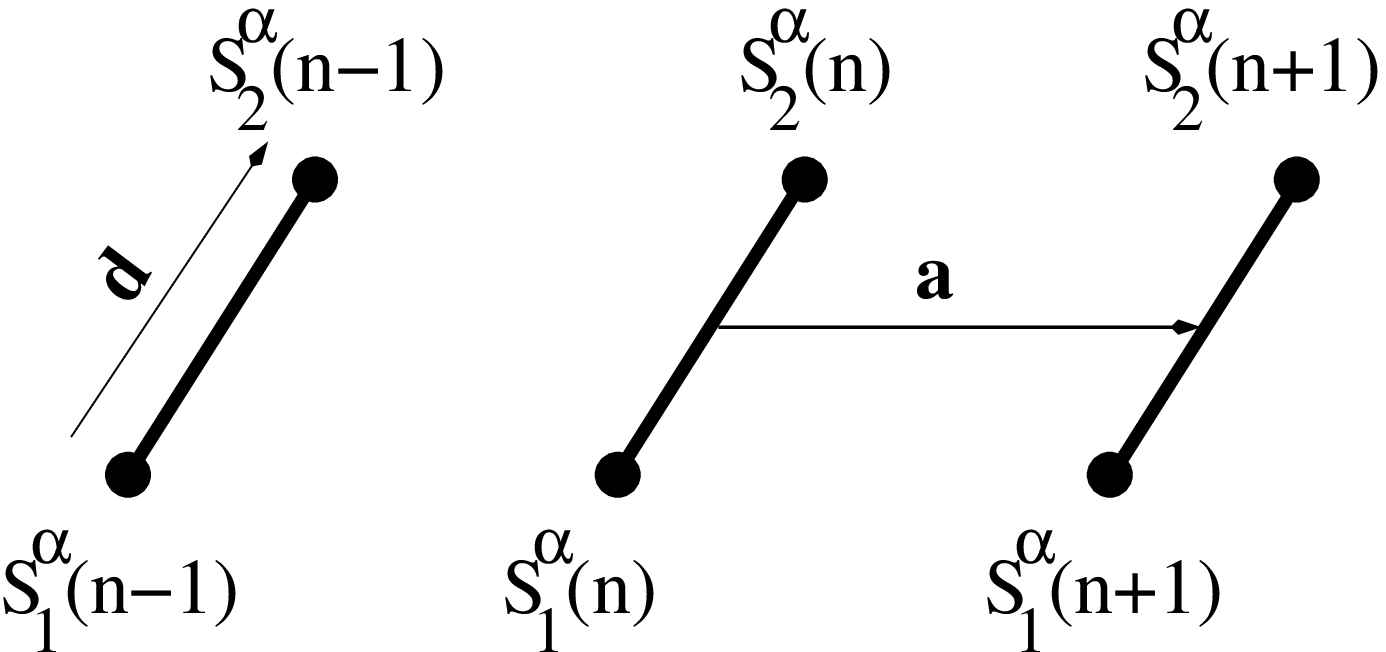}
\caption{\label{fig:dimerstruct}
Linear arrangement of dimers indicating the similarity of
alternating chain (interdimer interactions only between $S_2(n)$ and
$S_1(n+1)$) and ladder (general interdimer interactions) geometries}
\end{center}
\end{figure}
As shown in \Fref{fig:dimerstruct} the cristallographic unit cell
contains two spins. Thus the Fourier transform of the spin operator
splits into two parts and reads as:
\begin{equation}
S^{\alpha}(\vec{q}) = 
    \sum_n e^{-iq na}\left(e^{-i \frac{\vec{q}\vec{d}}{2}} S^{\alpha}_1(n)  
           + e^{i \frac{\vec{q}\vec{d}}{2}} S^{\alpha}_{2}(n)\right).
\end{equation}
As before $\|\vec{a}\| = a$ is the distance between neighbouring spins
and $\vec{d}$ denotes the spacing between the two spins on a
dimer. In our notation $q$ is the projection of the wave vector
$\vec{q}$ on the chain direction.

Using translational invariance with respect to the center of the dimer
we obtain for the singlet-triplet transition amplitude:
\begin{eqnarray}\label{eq:int}
I_{sm}(\vec{q}) = \sum_n e^{-iqan} 
\left[A^{zz}_{11}(n)+A^{zz}_{22}(n)+e^{i\vec{q}\vec{d}}A^{zz}_{12}(n) +
                   e^{-i\vec{q}\vec{d}}A^{zz}_{21}(n)\right]\\
\fl{\rm where} \qquad
A^{zz}_{ij}(n) = \bra 0 \mathcal S_i^{z}(0) \ket t \bra t  {\mathcal 
S_j^{z}}^{\dagger}(n) \ket 0, \qquad i,j=1,2.
\end{eqnarray}
Here, the sum is taken over all integer numbers $n$. However, it is more
convenient to calculate the functions $A^{zz}_{ij}(n)$ for positive
numbers $n$. This is feasible making use of inversion symmetry
wrt to the dimer center, implying
\begin{equation}\label{eq:inversion}
A^{zz}_{11}(-n) = A^{zz}_{22}(n)\quad{\rm and}\quad A^{zz}_{12}(-n) 
= A^{zz}_{21}(n).
\end{equation}
Inserting \eref{eq:inversion} into \eref{eq:int} one arrives at the
following result:
\begin{eqnarray}
I_{sm}(\vec q) & =  2 \sum_{n>0} \left(A^{zz}_{11}(n) + 
A^{zz}_{22}(n)\right) \cos(qna)\\
&+ 2 \sum_{n>0} \left(A^{zz}_{12}(n) 
\cos(qna-\vec{q}\vec{d})+A^{zz}_{21}(n) 
\cos(qna+\vec{q}\vec{d})\right)\\
&+ A^{zz}_{11}(0) + A^{zz}_{22}(0) + 
e^{i\vec{q}\vec{d}}A^{zz}_{12}(0)+e^{-i\vec{q}\vec{d}}A^{zz}_{21}(0).
\end{eqnarray}
Now, functions $A^{zz}_{ij}(n)$ enter for positive $n$ only. In the limit of
noninteracting dimers only the terms with $n=0$ (last line) survive.

At first glance the functions $A^{zz}_{ij}(n)$ are groundstate
expectation values which can be computed by the well established cluster
expansion method \cite{Singh90}. The projection operator $\mathcal P =
\ket t \bra t$ has to be evaluated from the one magnon states
$\ket{\psi^{(i)}}$, where $i$ labels the lattice site. By means of
degenerate cluster expansion these states are generated order by
order \cite{Gelfand95}. Then we find for projection operator:
\begin{equation}\label{eq:projection}
\mathcal P  = \ket 1 \bra 1 = \sum_{i j} (g^{-1})_{ij} 
\ket{\psi^{(i)}}\bra{\psi^{(j)}}.
\end{equation}
$g$ is the overlapping matrix of the $\ket{\psi^{(i)}}$:
\begin{equation}
g_{ij} = \langle\psi^{(i)}\ket{\psi^{(j)}}.
\end{equation}
To invert $g$ we use the fact that $g$ is the unit matrix for 
$\lambda\to 0$:
\begin{equation}
g = I + \tilde g.
\end{equation}
Owing to the matrix norm $\|\tilde{g}\| < 1$ we apply a geometric 
series to invert $g$:
\begin{equation}\label{eq:ginvert}
(I + \tilde g)^{-1} = \sum_{i=0}^{\infty} {\tilde g}^i.
\end{equation}
Now we have everything at hand to calculate the
singlet-triplet-intensity of the dynamical structure factor: Apply
degenerate perturbation theory to obtain the states $\ket{\psi^{(j)}}$
and $\mathcal P$. Then calculate $g$ and invert this matrix by using
\eref{eq:ginvert}. Finally, apply non degenerate perturbation theory
to compute the functions $A^{zz}_{ij}(n)$.

\ack
The experienced support of F. Demmel and A. Hiess during the IN3 
experiment is gratefully acknowledged. This work was partially 
supported by the Swiss National Science Foundation through the NCCR project MaNEP.

\end{document}